\documentclass[prl,twocolumn,showpacs,superscriptaddress,nobibnotes,amsmath,amssymb,nofootinbib,aps,10pt]{revtex4-2}

\usepackage{graphicx}
\usepackage{lineno}
\usepackage{dcolumn} 
\usepackage{bm}
\usepackage{hyperref}
\usepackage{textcomp}
\usepackage{textcomp}
\newcommand{\nocontentsline}[3]{}
\newcommand{\toclesslab}[3]
{\vspace{0.1in}\bgroup\let\addcontentsline=\nocontentsline#1{#2\label{#3}}\egroup\vspace{-0.1in}}
\usepackage{xcolor}
\usepackage[sectionbib]{bibunits}
\usepackage{mydef}
\usepackage{soul}
\usepackage{newfloat}
\DeclareFloatingEnvironment[
  fileext=lof,
  name=Extended Data Fig.,
  listname=List of Extended Data Figures
]{extendedfigure}
\begin{document}

\title{
The physical consequences of sperm gigantism}


\author{Jasmin Imran Alsous\textsuperscript{\dag}}
\affiliation{Center for Computational Biology, Flatiron Institute, New York, NY, USA}

\author{Brato Chakrabarti\textsuperscript{\dag,*}}
\affiliation{International Centre for Theoretical Sciences, Tata Institute of Fundamental Research, Bengaluru 560089, India}


\author{Bryce Palmer}
\affiliation{Center for Computational Biology, Flatiron Institute, New York, NY, USA}
\affiliation{Department of Mechanical Engineering, Michigan State University, East Lansing, MI 48864, USA}

\author{Michael J. Shelley\textsuperscript{*}}
\affiliation{Center for Computational Biology, Flatiron Institute, New York, NY, USA}
\affiliation{Courant Institute, New York University, New York, NY, USA}

\vspace*{-1em}
\begingroup
\renewcommand\thefootnote{\dag}
\footnotetext{Equal contribution}
\endgroup

\begingroup
\renewcommand\thefootnote{*}
\footnotetext{
\texttt{brato.chakrabarti@icts.res.in} \\ \texttt{mshelley@flatironinstitute.org}}
\endgroup


\begin{abstract}

The male fruit fly produces $\sim$1.8 mm long sperm, thousands of which can be stored until mating in a $\sim$$200 \ \mu$m sac, the seminal vesicle \cite{Lefevre1962, Tokuyasu1974}. While the evolutionary pressures driving such extreme sperm (flagellar) lengths have long been investigated \cite{Joly1995,Bjork2006,Manier2010, Pattarini2006, Miller2002}, the physical consequences of their gigantism are unstudied. Through high-resolution three-dimensional reconstructions of \textit{in vivo} sperm morphologies and rapid live imaging, we discovered that stored sperm are organized into a dense and highly aligned state. The packed flagella exhibit system-wide collective ‘material’ flows, with persistent and slow-moving topological defects; individual sperm, despite their extraordinary lengths, propagate rapidly through the flagellar material, moving in either direction along material director lines. To understand how these collective behaviors arise from the constituents' nonequilibrium dynamics,
we conceptualize the motion of individual sperm as topologically confined to a reptation-like tube formed by its neighbors. Therein, sperm propagate through observed amplitude-constrained and internally driven flagellar bending waves, pushing off counter-propagating neighbors. From this conception, we derive a continuum theory that produces an extensile material stress that can sustain an aligned flagellar material \cite{aditi2002hydrodynamic,saintillan2008instabilities1,gao2017analytical}. Experimental perturbations and simulations of active elastic filaments verify our theoretical predictions. 
Our findings suggest that active stresses in the flagellar material maintain the sperm in an unentangled, hence functional state, in both sexes, and establish giant sperm in their native habitat as a novel and physiologically relevant active matter system.

\end{abstract}

\maketitle

\noindent \textbf{Main} 

\vspace*{0.2em}

\begin{figure*}
    \centering
    \vspace*{-1em}
    \includegraphics[width=0.8\textwidth]{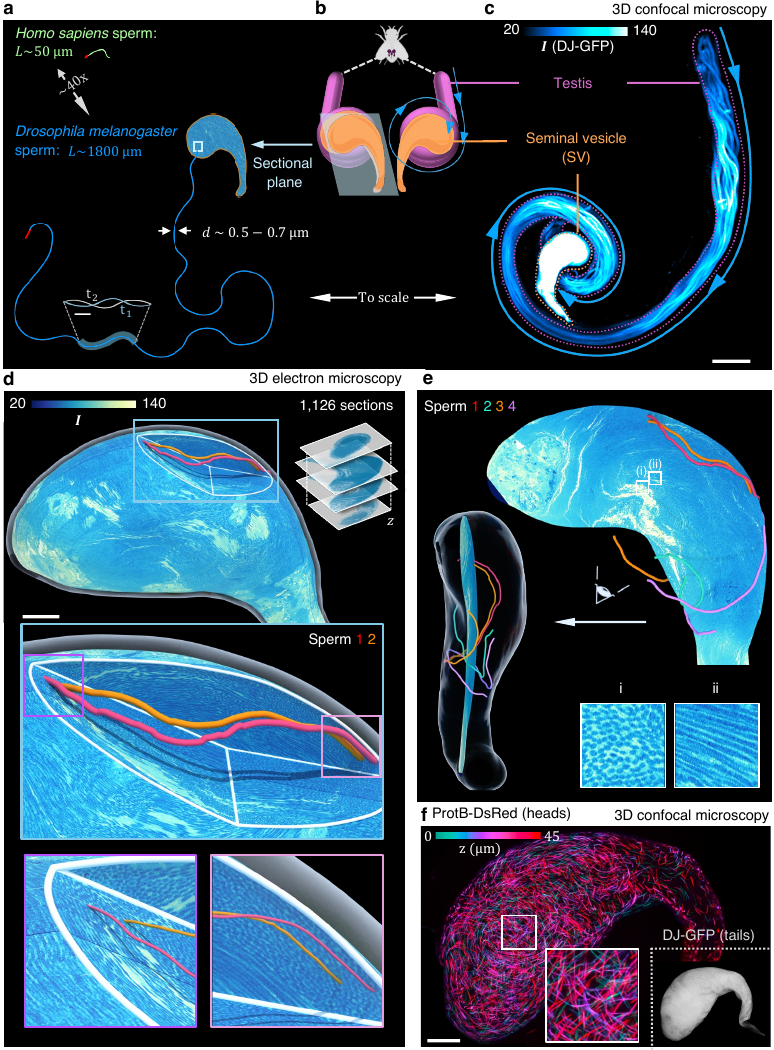}
    \vspace*{-3mm}
    \caption{\textbf{3D organization and packing of giant \textit{D. melanogaster} sperm in the seminal vesicle (SV).} \textbf{a,} 
    Schematic of a single \textit{D. melanogaster} sperm traced from a confocal image (Extended Data Fig.~\ref{fig:EDF1})\textemdash shown to scale relative to a \textit{Homo sapiens} sperm and to the male's sperm storage organ, the SV, in \textbf{b} and \textbf{c}. The sperm generates bending waves (Supplementary Video~1); inset shows two overlaid successive time points; scale bar = 10 $\mu$m. \textbf{b,} Simplified schematic of the male's reproductive system. Sperm produced in the $\sim$2 mm-long testes (magenta) are deposited in a pair of SVs (orange) (Supplementary Fig.~1). \textbf{c,} Maximum intensity projection (MIP) of fluorescently-labeled elongating sperm tails in the testis and mature sperm in the SV. Color indicates the intensity of the DJ-GFP signal, highlighting the relatively dense packing of sperm in the SV; scale bar = 100 $\mu$m. \textbf{d,} A 3D rendered image of all sperm in the SV, colored by intensity $I$, acquired through serial block-face scanning electron microscopy (SBF-SEM) of the entire storage organ (Extended Data Fig.~\ref{fig:EDF2}, Supplementary Video~2)\textemdash a portion of the data is digitally removed to expose the packing of sperm in the bulk; partial views of two segmented sperm (orange and magenta), spanning regions of distinct local alignment, are shown; scale bar = 20 $\mu$m. \textbf{e,} A single SBF-SEM section with four partially segmented sperm; also shown is an alternative viewpoint with the SV's outer boundary demarcated (white shell). Two regions (white boxes) are magnified to highlight the individual sperm's \textit{(i)} out of- and \textit{(ii)} in-plane packing in the SV, respectively.
    \textbf{f,} MIP of fluorescently-labeled sperm heads (ProtB-DsRed), color-coded by depth ($z$) in the SV (Supplementary Video~2), with one region magnified; scale bar = 20 $\mu$m. Rightmost inset is the corresponding MIP of the flagella (DJ-GFP).} 
    \label{fig:Fig1}
\end{figure*}

\noindent Some animals produce exceptionally long sperm, comparable to the animal itself, and much longer than the distance traveled in the female. Such `giant' sperm have evolved in several animal taxa \cite{Taylor1982, Soldatenko2018, Smith2016, Pitnick2009, Pitnick1995, Joly1995, Fitzpatrick2022} for $>$100 million years \cite{Wang2020}\textemdash a testament to the enduring success of sperm gigantism as a sexual strategy. 
Males of the fruit fly \textit{Drosophila melanogaster}\textemdash a powerful and relevant model organism for quantitative and mechanistic studies of reproduction \cite{Manier2010, Miller2002,Jackson2023, Snook2004, BlochQazi2003}, produce $\sim$$1800 \ \mu$m long sperm (Fig.~\ref{fig:Fig1}a, Extended Data Fig.~\ref{fig:EDF1}) \cite{Joly1989, Joly1994}. Most of the sperm's length is occupied by a relatively thick ($d \sim 0.5-0.7 \ \mu$m \cite{Laurinyecz2019,Noguchi2011}) and active flagellum that is powered by bending waves generated by the conserved microtubule-based axoneme (Supplementary Video~1) \cite{Perotti1973, Tokuyasu1974, Phillips1970,Mencarelli2008}; 
 the needle-shaped head is merely $\sim$$10 \ \mu$m long. 
 Sperm produced in the testes as bundles are deposited as individuals in two seminal vesicles (SVs) -- $L \sim 200 \ \mu$m sac-like organs where motile sperm remain until mating \cite{Lefevre1962, Kaufman1942, Tram1999} (Fig.~\ref{fig:Fig1}b,c, Supplementary Fig.~1).
 The dense packing of long thin objects under confinement is a ubiquitous biological phenomenon
 \textemdash perhaps its most notable biological realization being the packing of meters of chromosomal DNA within a $10 \ \mu$m human nucleus \cite{LiebermanAiden2009}. An important class of these problems arises when the filaments themselves are internally active; 
 one such example is the centimetric California blackworm, hundreds of which can form a tightly-wound three-dimensional (3D) tangled structure, but can rapidly disentangle under duress \cite{Patil2023}. Sperm in the SV must resist knots and tangles if the animal is to remain fertile \cite{Yang2011, Mojica1996}; how such sperm organize and how their autonomous activity manifests collectively within the tight confines of the reproductive system is unknown. 


Here, we report the discovery of an \textit{in vivo} dense active matter system, whereby thousands of giant sperm form a highly aligned and structured motile state, punctured only by spontaneous and persistent topological defects, resembling a living liquid crystal \cite{sanchez2012spontaneous,Zhou2014}. 
Highly resolved and rapid microscopy, perturbation experiments and simulations of active elastic filaments demonstrate that these dynamics arise from contact interactions between counter-propagating flagella that actively reptate within the flagellar mass. A minimal and experimentally constrained coarse-grained theory built on this conception rationalizes the sperm's ordered motility, separation of time scales, and system-wide material flows in the SV, as well as the observed \textit{in vivo} dynamics in the female's dedicated storage organs. Our work suggests a role for active material stresses in fluidizing sperm in their natural habitat, with implications for the animal's fertility.


\vspace*{0.2em}

\noindent \textbf{3D \textit{in vivo} organization of giant sperm}

\vspace*{0.2em}

\noindent We first set out to reconstruct the 3D organization and packing of sperm in the SV, readily extractable from the male's abdomen through dissection (Fig.~\ref{fig:Fig1}\textbf{a,b}, Supplementary Section I). Using serial block-face scanning electron microscopy (SBF-SEM), we acquired volumetric images spanning the entire SV and its resident sperm. By revealing the nanoscale ultrastructure and anatomical features of the entire SV \cite{Denk2004}, SBF-SEM allows us to resolve all individual sperm in the organ (Supplementary Video~2, Supplementary Section I). We discovered that the SV ($\approx 200 \ \mu\mathrm{m} \times 150 \ \mu\mathrm{m} \times 150 \ \mu\mathrm{m}$), independently of the sectional orientation (Extended Data Fig.~\ref{fig:EDF2}), is densely packed with sperm whose alignment is correlated over a length scale of $\sim$$48 \pm 12 \ \mu$m (Supplementary Section II). Reconstructions of segmented individual sperm reveal that a sperm's flagellum meanders through the SV, with low-amplitude periodic undulations along its backbone presumably resulting from the internally generated deformation waves (Fig.~\ref{fig:Fig1}\textbf{d}; middle panel). The sperm's length being longer than the correlation length of director orientation suggests that it transits through multiple regions of nematic orientation (Fig.~\ref{fig:Fig1}\textbf{e}). 

Next, we performed 3D confocal live imaging of sperm in the SV whose flagella and heads were illuminated through fluorescent labeling of Don-Juan (DJ-GFP), a male germline-specific mitochondrial protein found along the entire length of the $\sim$$1800 \ \mu$m flagellum \cite{Santel1998}, 
and ProtamineB (ProtB-DsRed), a protein involved in compacting sperm DNA \cite{JayaramaiahRaja2005}, 
respectively (Supplementary Section I). Consistent with our findings from SBF-SEM, we found that sperm in the SV are packed into a dense, aligned state\textemdash evident from examining the spatial organization of both their relatively short heads and much longer flagella (Fig.~\ref{fig:Fig1}\textbf{f}). We estimate that sperm volume fraction $\varphi$ in the SV ranges from $\sim$$40-60\ \%$ (Supplementary Section II).

\begin{figure*}
    \centering
    \includegraphics[width=0.9\textwidth]{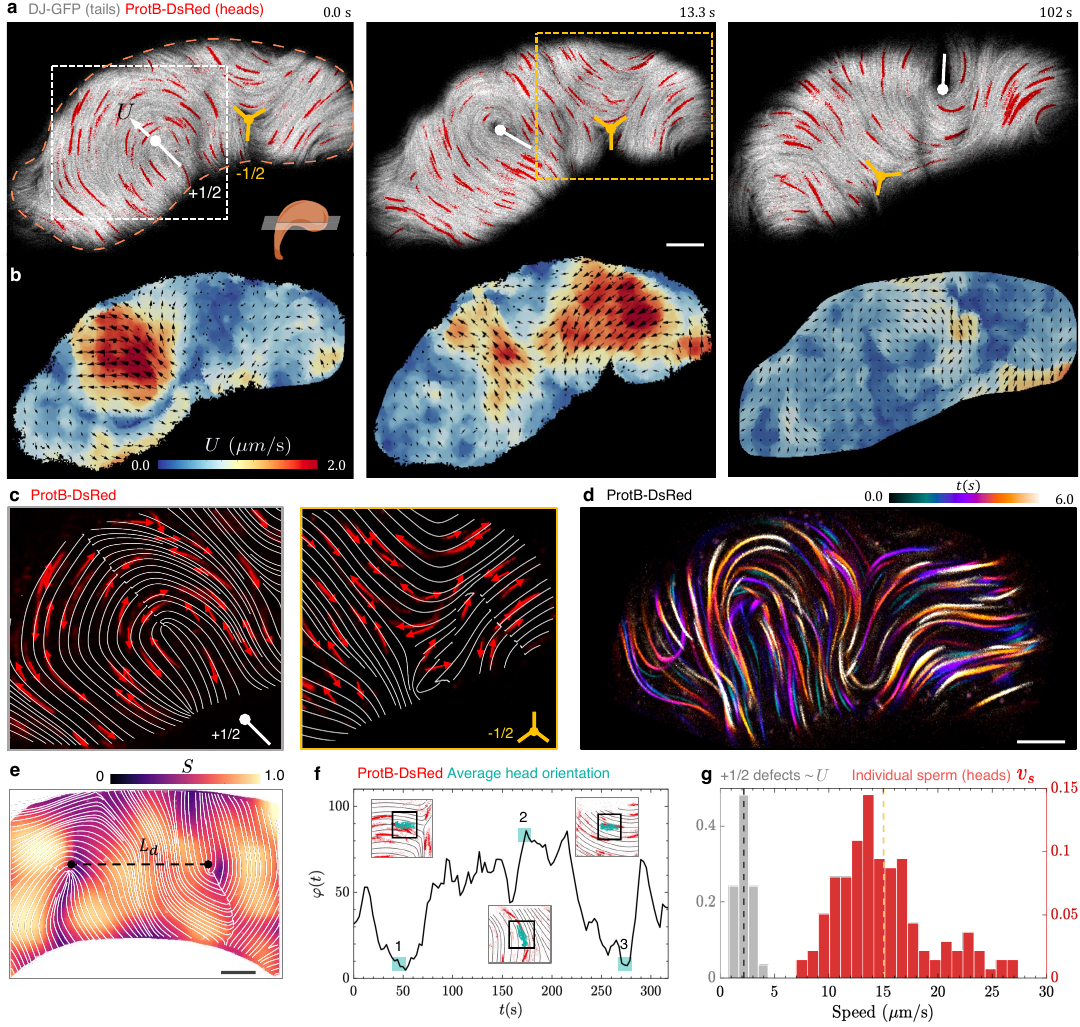}
    \caption{\textbf{Spatiotemporal dynamics of individual giant sperm and their \textit{in vivo} collective flows.} \textbf{a,} Snapshots from a time-lapse of fluorescently labeled sperm (flagella: white, DJ-GFP; heads: red, ProtB-DsRed) in an optical section (depth $\sim0.25 \ \mu$m) through the SV (Supplementary Video 2). The flagella exhibit continuously evolving system-wide `material' flows, characterized by alignment and persistent $\pm 1/2$-order topological defect dynamics. Boxed regions are magnified in \textbf{c}. Orange dashes delineate the boundary of the SV (Supplementary Video~2). \textbf{b,} Material velocity fields corresponding to \textbf{a}, with the magnitude of $U$ color-coded; arrows are velocity vectors. Measurements of $U \sim 2 \ \mu$m/s through optic flow are in agreement with estimates obtained by measuring the timescale of evolution of the local average director orientation (see \textbf{f}) and from tracking the motion of $+1/2$-order defects (see \textbf{g}). \textbf{c,} Sperm heads (ProtB-DsRed, red) with the nematic director field overlaid in proximity to a (\textit{left}) $+1/2$- and a (\textit{right}) $-1/2$-order defect. Sperm swim along the nematic director field lines with arrows indicating their direction of motion. Individual sperm move more rapidly ($v_s \sim 15 \ \mu$m/s) than the emergent flagellar material flows (see \textbf{g}). \textbf{d,} Color-coded maximum intensity projection (MIP) of successive frames from a 6-second window from a time-lapse of a SV with fluorescently labeled sperm heads (ProtB-DsRed), capturing the sperm's laning behavior. \textbf{e,} Snapshot of a SV from a time-lapse video showing the separation between a pair of $+1/2$-order defects (average $L_d \sim 70 \pm 11 \ \mu$m, $n$ = 12 $+1/2$-order defect pairs). The background color indicates the nematic order parameter $0\leq S \leq 1$, highlighting the large-scale alignment of sperm. Scale bars in \textbf{a}, \textbf{d}, and \textbf{e} = 10 $\mu$m. \textbf{f,} Average orientation of sperm heads in an ROI in an optical slice as a function of time (Supplementary Section II). Insets show the heads' (ProtB-DsRed, red) average orientation (cyan)  within that ROI, with the overlaid director field. \textbf{g,} Histograms of individual sperm (red) and $+1/2$-order defect (gray) speeds in the SV, highlighting a separation of time scales between their respective dynamics; dashed lines indicate the respective averages ($n$ = 174 individual sperm in 9 SVs; $n$ = 41 +1/2-order defects in 10 SVs).}
    \label{fig:Fig2}
\end{figure*}

\vspace*{0.5em}
\noindent \textbf{Collective material flows and individual laning}
\vspace*{0.2em}

\noindent We next turn to the sperm's individual and collective dynamics. We suggest that the reader first watch Supplementary Video~3—a live imaging time-lapse that captures sperm motion, with separately labeled heads and tails, in their dense \textit{in vivo} environment. These data reveal that despite their dense packing, sperm in the SV are highly motile. 
Specifically, in contrast to the seemingly passive, and bundled sperm in the testes \cite{Lefevre1962} the sperm's flagella in the SV collectively fold together, resulting in continuously evolving and coherent system-wide flagellar flows which can last for hours following extraction of the SV (Fig.~\ref{fig:Fig2}; Supplementary video~3). We also observe persistent $\pm 1/2$-order topological defects (Fig.~\ref{fig:Fig2}\textbf{a-c}), cores devoid of sperm, spontaneously formed by the folding of long flagella\textemdash similar to those exhibited by deforming fibers in cavities \cite{Donato2002} and growing long-chain bacterial colonies \cite{Faluweki2023}. Live imaging shows that flagella are highly aligned, as characterized by the scalar order parameter $S$. The overlaid nematic director field further reveals that changes in flagella orientation occur over a characteristic length scale of $\sim$$70 \ \mu$m, given by the separation distance of pairs of $+1/2$-order defects (Fig.~\ref{fig:Fig2}\textbf{e}), consistent with our data from fixed samples (Fig.~\ref{fig:Fig1}).
 
By tracking the moving comet-shaped $+1/2$-order topological defects, we obtained a proxy value for the material speed $U \sim 2.5 \pm 0.9 ~\mu$m/s; $-1/2$-order defects are, in contrast, much slower with speeds $\sim$$0.67 \ \pm  0.26 \mu$m/s (Fig.~\ref{fig:Fig2}\textbf{g}; Supplementary Section II). The direction of motion of the $+1/2$-order defects (rear-to-front) indicates that the flagellar material is under an extensile (\textleftarrow \ \textrightarrow), rather than a contractile (\textrightarrow \ \textleftarrow), stress along the material director field \cite{copenhagen2021topological}. The director field orientation changes with a characteristic time scale $\tau \sim 45 \pm 15$s (Fig.~\ref{fig:Fig2}\textbf{f})\textemdash interpretable as a measure of the inverse time associated with the mean vorticity of the material flow (Supplementary Section II). Given a typical system size $L \sim 200 \mu$m, we find that $U \sim L/\tau \sim 3-4 \mu$m/s\textemdash in good agreement with that obtained from measuring the speeds of $+1/2$-order defects. From the same microscopy data, we also tracked the sperm's needle-shaped heads: being the same width as, yet much shorter than the tail (Fig.~\ref{fig:Fig1}\textbf{a}; Extended Data Fig.~\ref{fig:EDF1}), the $\sim$$10 \mu$m long heads serve as natural, sparser markers for sperm motion in the dense flagellar material. We found that sperm swim relatively quickly ($v_s \sim 15.6 \pm 5.1 \mu$m/s) along the nematic director lines (Fig.~\ref{fig:Fig2}\textbf{g}). When successive frames in a time-lapse are superimposed, the sperm's laning behavior can be captured in a still (Fig.~\ref{fig:Fig2}\textbf{d}), just as long-exposure photography captures the laning of cars on highways. 
Upon compressing the SV under a coverslip, thus confining the flagellar flows to a quasi-2D geometry (Supplementary Section I), we observe system-wide material flows with qualitatively similar features to those in 3D (Fig.~\ref{fig:Fig5}\textbf{a,b}, Extended Data Fig.~\ref{fig:EDF4}, Supplementary Video~5).

\vspace*{0.5em}
\noindent \textbf{Photomanipulation reveals apolar organization}
\vspace*{0.2em}

\noindent As a first step towards understanding the mechanism underlying the sperm's fast and smooth laning and slow collective material flows, we characterized the sperm's relative motion in the SV. Again using time-lapses, we identified and examined 948 contiguous pairs of heads (separated by less than $\sim$$0.5 \ \mu$m ; Supplementary Section II) and found that $\sim$$80$\% of those pairs comprised sperm heads that swam in opposite directions (Fig.~\ref{fig:Fig3}\textbf{a,b}). To relate this to the behavior of their flagella, we selectively photobleached regions of fluorescently labeled active flagellar material (director field intensity and individual sperm speeds remain largely unaltered pre- and post-photobleaching; Supplementary Section II). As photobleaching marks the subset of the flagella in the region of interest (ROI), such an experiment uncovers the relative dynamics of the individual constituents, as has been effectively done for microtubules in active motor-driven {\it in vitro}  assemblies \cite{sanchez2012spontaneous,Frthauer2019}. 

Photobleaching transversely to the nematic field revealed that adjacent flagellar segments moved predominantly in opposite directions along the field lines (Fig.~\ref{fig:Fig3}\textbf{c-f}; Supplementary Video~4), consistent with results obtained from tracking the head pairs. 
The extent of interdigitation of bleached sperm segments can be measured by $\Lambda$, defined as the distance between two bleached segments moving in the same direction when separated by a bleached segment(s) moving in the opposite direction (Supplementary Section II). As such, $\Lambda$ has a lower bound of one sperm diameter ($d \sim 0.5-0.7 \ \mu$m \cite{Laurinyecz2019,Noguchi2011}), realized by a completely interdigitated arrangement of counter-propagating sperm that constitute a locally apolar material. Conversely, higher values of $\Lambda$ reflect the unidirectional co-propagation of adjacent sperm, consistent with a region of higher polarity (Fig.~\ref{fig:Fig3}\textbf{e,f}). Our measurements reveal a clustering of the data around values of $\Lambda$ comparable to the diameter of $\sim$$1-2$ sperm, with $\bar{\Lambda} = 1.8 \pm 1.2$ (Fig.~\ref{fig:Fig3}\textbf{h}). These data suggest that the relative motion of sperm with respect to each other is generally that of counter-propagation. These results are consistent with data from photobleaching experiments in a quasi-2D geometry (Supplementary Section II, Supplementary Video~5). We further observed that such co-propagating bleached segments moved more slowly, compared to their interdigitated counterparts. Notably, repetitive photobleaching of polar regions also revealed the local evolution of the sperm's relative motion, whereby polar regions became progressively apolar (Extended Data Fig.~\ref{fig:EDF3}, Supplementary Video~4)\textemdash possibly driven by the ability of {\textit{Drosophila}} and other \textit{Diptera} sperm to propagate bending waves from tip-to-base or from base-to-tip (Supplementary Video~1) \cite{Yang2011, Mojica1996}. Although our analysis focused on regions of uniformly aligned sperm in the SV, we observed similar interdigitation of bleached flagella along director lines at topologically more complex regions like a tri-junction (Fig.~\ref{fig:Fig3}\textbf{g}, Supplementary Video~4). Taken together, our quantitative analyses of the sperm's \textit{in vivo} dynamics reveal a largely nematic material with predominantly counter-propagating flagella.

\begin{figure*}
    \centering 
    \label{fig:Fig3}
    \includegraphics[width=0.9\textwidth]{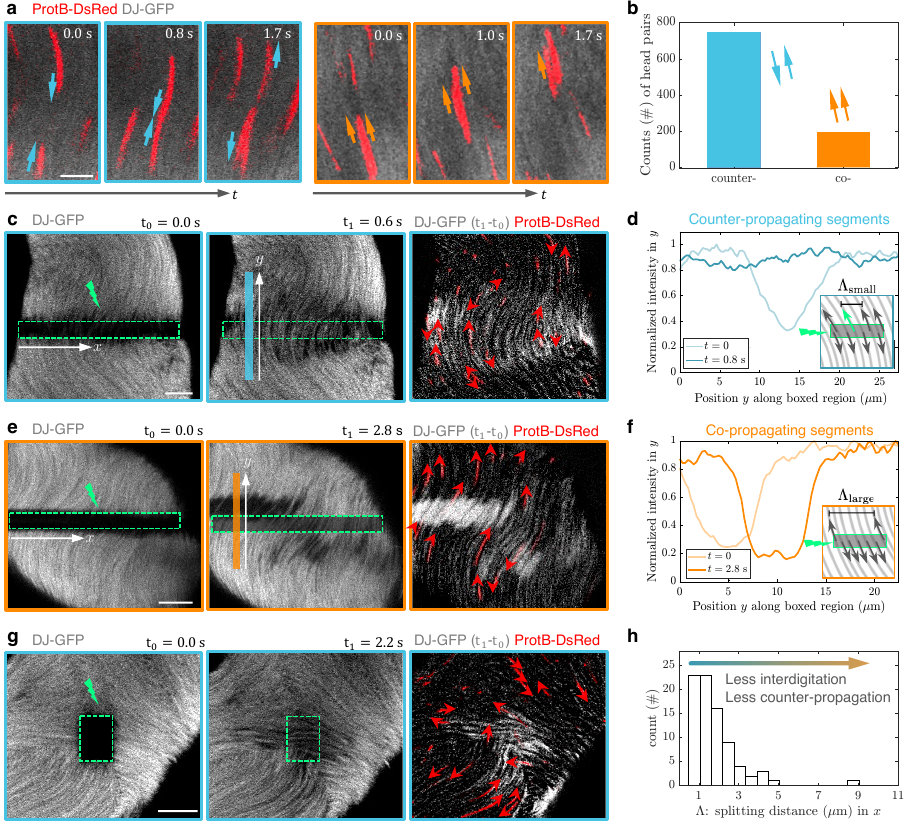}
    \caption{\textbf{Photomanipulation experiments reveal the sperm's nematic and largely apolar arrangement}. \textbf{a,} Snapshots from time-lapses of sperm in the SV with fluorescently labeled heads (ProtB-DsRed, red) and tails (DJ-GFP, gray). Two contiguous pairs of heads are shown, with blue and orange arrows indicating the heads' relative direction of motion: counter- and co-propagating, respectively (scale bar = 5 $\mu$m). \textbf{b,} Bar chart of the number of counter- and co-propagating contiguous head pairs in SVs ($n$ = 948 pairs, Supplementary Section II). \textbf{c,} Snapshots from a time-lapse of fluorescently labeled sperm in the SV, during which a rectangular region of interest (ROI, outlined in green), perpendicular to the sperm’s nematic director, was photobleached at $t=0.0$ s. Photobleaching marks the sperm within the photobleached ROI, revealing the relative motion of a subset of flagella within this dense material. Adjacent photobleached flagellar segments counter-propagate along the nematic field lines at $t=0.6$ s. The third panel shows the subtraction of the DJ-GFP signal in the two time frames to aid in visualization, with sperm heads (ProtB-DsRed) overlaid (arrows indicate their direction of motion.) \textbf{d,} Normalized intensity as a function of distance in $y$ along the region in blue in \textbf{c} at $t=0.0$ s and at $t=0.6$ s, demonstrating the rapid recovery of fluorescence through the influx of unbleached and interdigitated sperm from above and below the bleached region. \textbf{e,} Snapshots from a photobleaching experiment showing the splitting of the bleached region into relatively larger clusters of co-propagating photobleached flagella. The third panel shows the subtraction of the DJ-GFP signal in the two time frames, with overlaid sperm heads. \textbf{f,} Normalized intensity as a function of distance along $y$ of the region in orange in \textbf{e,} at $t = 2.8$ s. The time-stamp highlights a slower recovery of fluorescence through the influx of co-propagating flagella when compared to the interdigitated splitting (\textbf{c,d}) in apolar regions. \textbf{g,} Snapshots from an experiment in which a rectangular ROI was photobleached at a topologically complex region resembling a `tri-junction', displaying qualitatively similar behavior as in \textbf{c}\textemdash namely, the interdigitated splitting of photobleached segments along director field lines; the third panel shows the subtraction of the DJ-GFP signal from the two time frames. \textbf{h,} Histogram of  the splitting distance $\Lambda$\textemdash a measurable parameter that reflects the material's polarity, given by the distance between alternating photobleached segments traveling in the same direction, separated by single or multiple bleached segments traveling in the opposite direction ($\bar{\Lambda} = 1.8 \pm 1.2$, $n=82$ from 23 photobleaching events in 15 SVs).}
\label{fig:Fig3}
\end{figure*}

In contrast to the persistent and directional motility of the sperm in the SV (Fig.~\ref{fig:Fig2}\textbf{a}, Supplementary Video~3) we found that sperm released from the SV into the surrounding buffer remain largely stationary despite persistent waves of deformation moving along their flagellum (Supplementary Video~1). The lack of progressive motility of isolated \textit{Drosophila} sperm \textit{in vitro} \cite{Bressac1991} and giant sperm in other species \cite{Tokuyasu1974, Werner2007, Mencarelli2008} contrasts with that of self-propelled cells, such as mammalian sperm or bacteria. These data suggest that spatial restriction experienced \textit{in vivo} underpins the directed motility of sperm. 

\vspace*{0.5em}
\noindent \textbf{Active reptation explains observed dynamics}
\vspace*{0.2em}

\begin{figure*}
    \centering
    \includegraphics[width=0.9\textwidth]{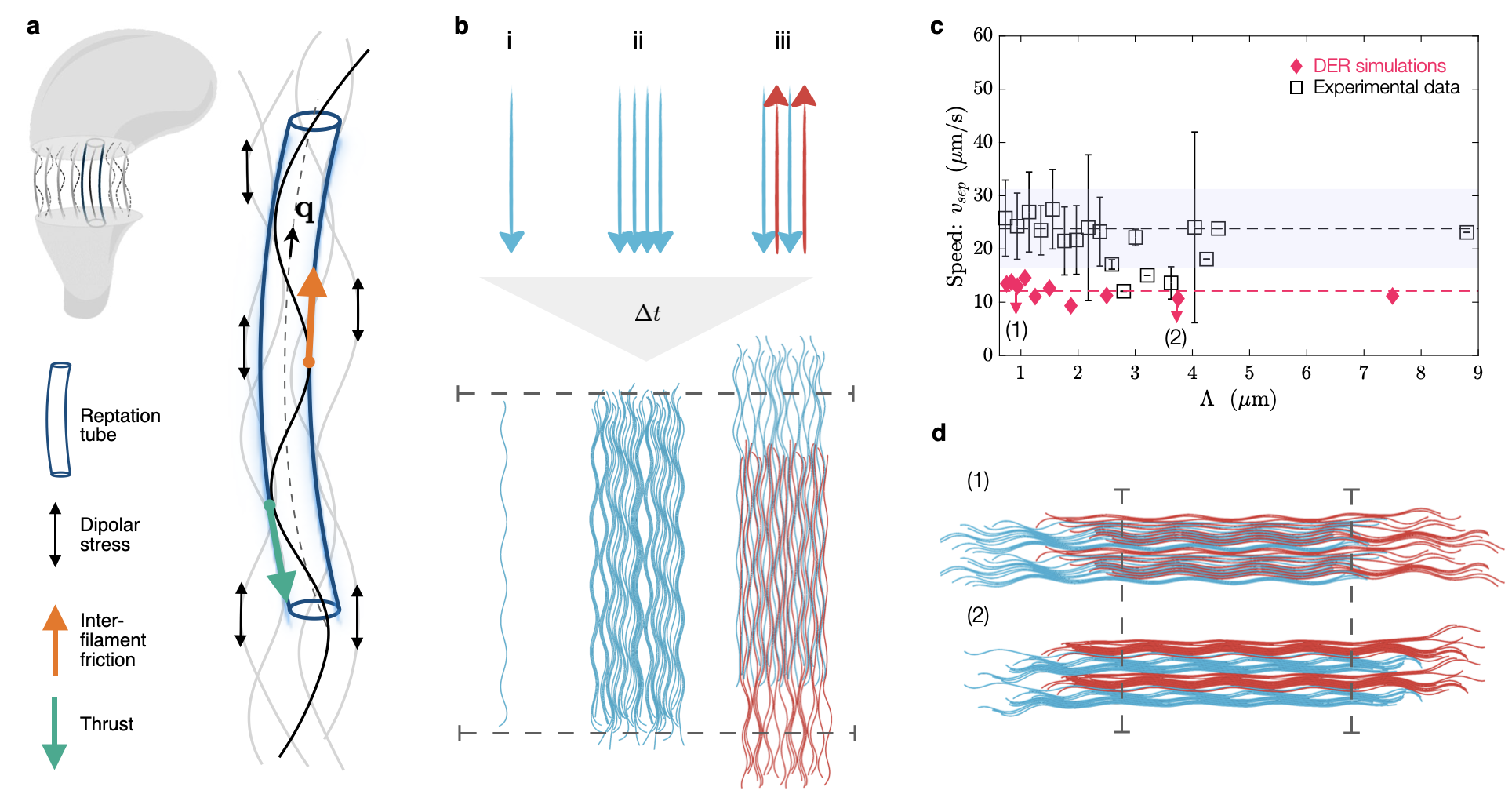}
\caption{\textbf{A theoretical model of reptating active filaments is supported by DER simulations and experimental perturbations.} \textbf{a,} Simplified schematic of a bisected SV showing transverse flagellar backbones of adjacent sperm and their bending waves (dashed lines). Each flagellum is constrained within a reptation tube (navy) formed by neighboring sperm (gray). Bending waves propagate along rod-like flagellar segments with orientation $\bq$, with an amplitude comparable to the radius of the confining tube. These waves interact mechanically with those of the adjacent sperm, giving rise to propulsive thrust (cyan arrow) and inter-filament friction (orange arrow). \textbf{b,}
Snapshots from DER simulations of active filaments for three different configurations (Supplementary Section V, Supplementary Video~6). The filaments' initial configurations are schematized on the top row with arrows indicating the direction of wave propagation (-$\bq$ in the hydrodynamic model). The evolved configuration at a later time is depicted in the bottom row, with the dashed lines indicating the initial positions of the filaments' end points. \textit{(i)}: An isolated active filament simulated with isotropic drag from the surrounding medium. In the absence of anisotropic friction, the filament is immotile. \textit{(ii):} An arrangement of horizontally periodic active filaments propagating bending waves in the same direction. In this polar configuration, the filaments become phase synchronized and steric interactions do not lead to directed motility. \textit{(iii):} A horizontally periodic interdigitated configuration where neighboring filaments propagate bending waves in opposite directions. In this apolar arrangement, steric interactions lead to the counter-propagation of active filaments, as observed in photobleaching experiments of apolar regions in the SV (Fig.~\ref{fig:Fig3}\textbf{c}). \textbf{c,} The separation speed $v_{\text{sep}}$ of bleached flagellar segments remains roughly constant as a function of the splitting distance $\Lambda$, as measured from photobleaching experiments ($n=82$ from 23 photobleaching events in 15 SVs; values of $\Lambda$ grouped into 40 bins ranging from 0.62-8.9 $\mu$m; Fig.~\ref{fig:Fig3}\textbf{h}). Overlaid are measurements of $v_{\text{sep}}$ from DER simulations. Data from experiments and DER simulations show good agreement with the prediction from the continuum model ($v_\text{sep} = |\dot{\bX}(\uvc{s})-\dot{\bX}(-\uvc{s})| \sim \text{constant}$). \textbf{d,} Arrangements of interlaced active filaments corresponding to two data points, (1) and (2), for DER simulations, with the initial position of the endpoints of the filaments marked by dashed lines (Extended Data Fig.~\ref{fig:EDF6}, Supplementary Video~8).}
\label{fig:Fig4}
\end{figure*}

\begin{figure*}
    \centering 
    \includegraphics[width=0.8\textwidth]{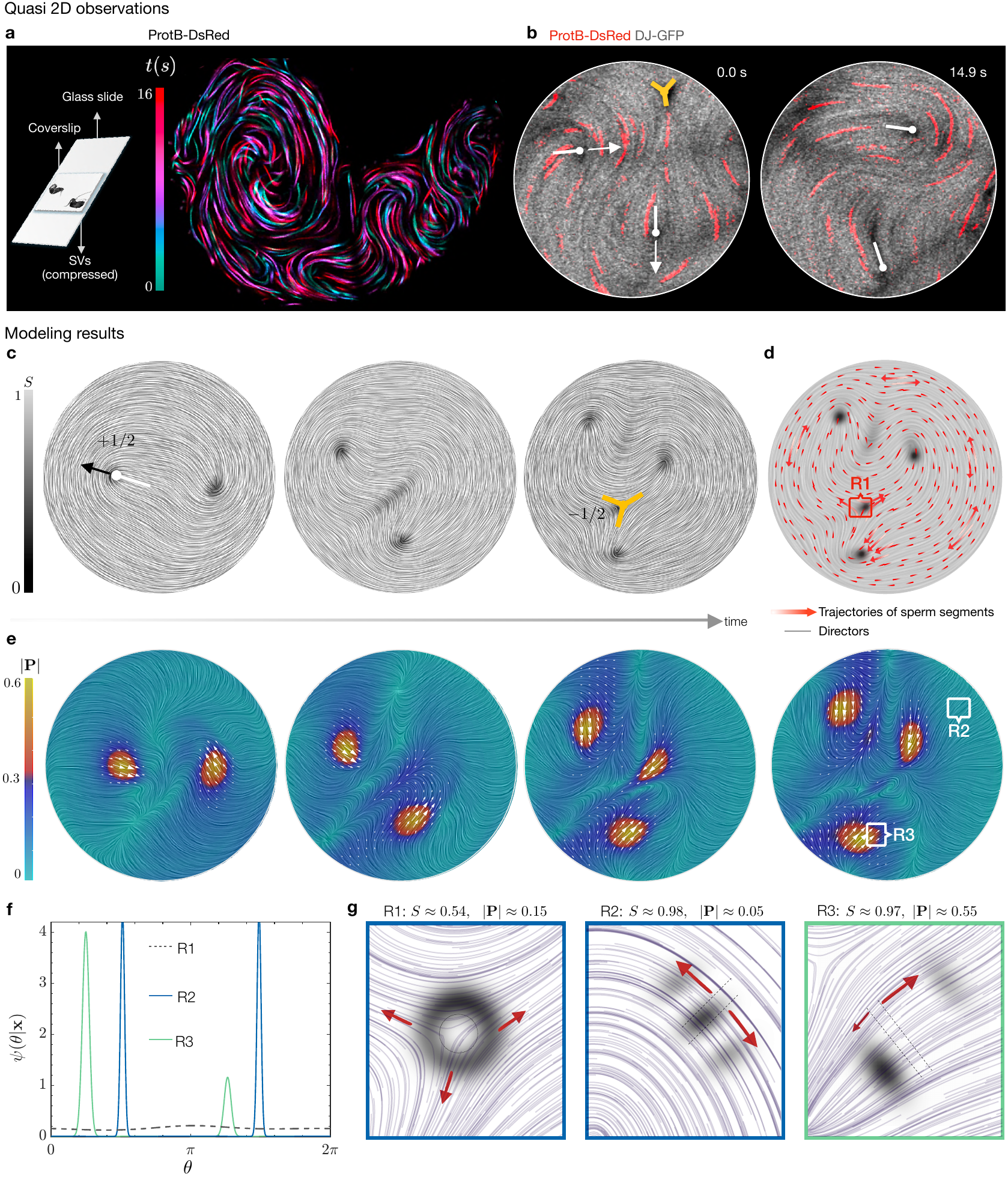}
    \caption{\textbf{Simulations of the active reptation model reproduce experimentally observed spatiotemporal dynamics.} \textbf{a,} 
    \textit{Left}: Schematic illustrating the compression of the SV between a coverslip and a glass slide, a setup that confines the flagellar flows to a quasi-2D geometry (Extended Data Fig.~\ref{fig:EDF4}, Supplementary Video~5). \textit{Right}: Color-coded maximum intensity projection (MIP) of successive frames from a time-lapse of a SV with fluorescently labeled sperm heads (ProtB-DsRed), demonstrating the sperm's laning behavior in 2D, as observed in 3D (Fig.~\ref{fig:Fig2}\textbf{d}). \textbf{b,} Live imaging of flagellar flows (DJ-GFP, gray) under a coverslip reveals $\pm 1/2$-order topological defects that emerge spontaneously through the subsequent folding of the flagellar material, as also observed in 3D (Fig.~\ref{fig:Fig2}\textbf{a}). \textbf{c,} From simulations in a disk, snapshots of the evolving director field of the flagellar material at successive time points, displayed using Line Integral Convolution (LIC); color encodes the the nematic order parameter $S$. Note the continuous bending of director fields and the spontaneous emergence of $\pm 1/2$-order topological defects (Supplementary Video~9). Simulations were initiated with small perturbations to the $\bQ$ tensor field around an aligned state; the emergent features are independent of the initial and boundary conditions (Supplementary Section IV). \textbf{d,} Trajectories of sperm segments (red) overlaid on the director field at a later time point, inferred from reconstructing approximations to the local distribution function $\psi(\bx,\bq)$ (see \textbf{f}). Sperm segments in apolar ($|\bP|\approx 0$) and nematic region counter-propagate along the directors, while those in polar regions ($|\bP|\gtrsim 0.5$) move unidirectionally in clusters. \textbf{e,} Snapshots of the associated polarity field $\bP$, where color encodes the strength of the local polarity $|\bP|$. Overlaid white arrowheads highlight the average direction of polar regions. \textbf{f,} Reconstruction of the microscopic organization of the flagellar material through reconstructing $\psi(\bx,\bq,t)$ (Supplementary Section IV), where $\bq = (\cos \theta, \sin \theta)$. $\psi(\theta|\bx)$ at a $-1/2$-order defect (R1), in a nematic apolar region (R2), and a nematic polar region (R3). \textbf{g,} Results from a numerical bleaching experiment on the simulated material. The bleached region is encapsulated by black dashed boundaries; the black bands highlight the subsequent positions of the bleached regions and their intensity represents the density of the bleached sperm segments. Arrows point to the direction of motion of the bleached regions; their lengths indicate their relative speeds. 
    }
 

   \label{fig:Fig5}
\end{figure*}

\noindent How might we make sense of our experimental observations? Given the sperm's dense packing and high alignment in the SV, we conceived of a theoretical model wherein an individual sperm's dynamics are geometrically constrained to move in an effective tube formed by its neighbors (Fig.~\ref{fig:Fig4}\textbf{a}). This is analogous to the reptation description of linear polymer chains in a melt wherein thermally-driven transverse fluctuations along the polymer backbone are restricted to an imaginary tube  \cite{de1971reptation}. A geometric scaling argument based on an idealized uniform packing of aligned sperm predicts that above a volume fraction of $\sim$$30$\%, inter-flagellar spacing is comparable to the amplitude of the bending waves (Supplementary Section III); given an average volume fraction of $\varphi 
\approx 47\%$ in the SV and the sperm's nematic alignment, wave propagation will result in contact-based interactions between adjacent flagellar backbones (Fig.~\ref{fig:Fig4}). The dense packing of sperm hinders direct \textit{in vivo} observation of moving flagellar waves (Supplementary Video~3); nonetheless, the deformation waves found in reconstructed sperm morphologies (Fig.~\ref{fig:Fig1}\textbf{d}), phase-synchronized waves  observed among co-propagating sperm within the SV (Supplementary Video~6), and the traveling bending waves moving along sperm released from the SV (Supplementary Video~1), support the thesis of contact interactions driven by flagellar waves \textit{in vivo}. While a bacterium derives its thrust and drag from the surrounding fluid, we hypothesize that SV sperm motility stems from steric and lubrication interactions between flagellar bending waves propagating in opposing directions. Given the sperm's dense packing in the SV, hydrodynamic interactions are heavily screened \cite{muthukumar1982screening}, with the surrounding fluid providing only hydrodynamic drag (Supplementary Section IV). 

To illustrate that contact forces between counter-propagating flagellar waves are sufficient to give rise to directed motility, we performed simulations of interacting discrete elastic rods (DER) simulations (Fig.~\ref{fig:Fig4}\textbf{b}; Supplementary Section V) \cite{bergou2008discrete,vetter2015packing}. Sperm are modeled as weakly-extensible slender filaments with aspect ratio 1:300 and a bending rigidity of a flagellum \cite{rikmenspoel_bull_sperm_1984}. Deformation waves moving along the elastic filament result from differences between the current and a prescribed time-periodic curvature wave in the elastic energy.
The overlap between adjacent filaments is prevented through excluded volume interactions, modeled through soft repulsive potentials. In Stokes flow, the propulsion of an isolated, undulating active filament results from the anisotropic drag forces experienced by the slender object \cite{gray1955propulsion}. To preclude this mechanism of propulsion, we impose an isotropic passive drag from the suspending fluid (Supplementary Section V); as a result, an isolated active filament fails to propel (Fig.~\ref{fig:Fig4}\textbf{b}(i), Supplementary Video~6). To investigate how contact interactions alter these dynamics in a dense material of such active and aligned filaments, we consider assemblies in two distinct arrangements. First, when the assembly is polar with all filaments propagating waves in the same direction (Fig.~\ref{fig:Fig4}\textbf{b}(ii)), the assembly remained largely stationary (Supplementary Video~6). Conversely, when the assembly is interdigitated and apolar, with neighboring filaments propagating bending waves in opposite directions (Fig.~\ref{fig:Fig4}\textbf{b}(iii)), the two polar sub-assemblies translate opposite to the their direction of wave propagation, resembling the experimentally observed splitting of bleached flagella in apolar regions of the SV (Fig.~\ref{fig:Fig3}\textbf{c}). Taken together, these simulations demonstrate that the reciprocal pushing of active filaments can give rise to their oppositely directed propulsion\textemdash a mechanism that differs from that of fluid-mediated thrust that drives the motility of mammalian or sea-urchin sperm \cite{gray1955propulsion,li2023chemomechanical}.

This conception of `moving by reciprocal pushing' leads naturally to a micro-macro continuum framework, with material stresses, that recapitulate the emergence of material flows in the SV. Akin to kinetic theory of polymeric liquids, we describe the flagellar conformation using a distribution function $\psi(\bx,\bq,t)$, the probability density of finding a flagellar segment with propulsive orientation $\bq \in \mathbb{S}^2$ (the unit sphere) at location $\bx$ \cite{doi1988theory}; bending waves on a flagellar segment propagate backwards along the propulsive orientation vector $\bq$. The flagellar material density is $\rho(\bx) = \int_{\mathbb{S}^2}\psi \md \bq$, and the local flagellar (or propulsive) polarity vector is $\mathbf{P}(\bx) = \rho^{-1} \int_{\mathbb{S}^2} \bq \psi(\bx)\md\bq$ (with scalar order $|\mathbf{P}|\in [0,1])$. The Fokker-Planck equation that evolves $\psi(\bx,\bq,t)$ depends on: (1) the sperm's translational velocity given by $\dot{\bx}(\bx,\bq,t) = \bU(\bx,t) + \dot{\bX}(\bx,\bq,t)$, where $\bU$ is the material flow velocity common to all flagellar segments at $\bx$ \cite{doi1988theory}, and $\dot{\bX} = v_s [\mathbf{P} - \bq]$ is the propulsion (`reptation') speed of individual segments ($v_s$ is a characteristic propulsion speed; see derivation in Supplementary Section IV); and (2) an orientational velocity $\dot{\bq}$  that is modeled using Jeffery's equation \cite{gao2017analytical}\textemdash a kinematic consequence of the flagellar backbones being rotated by the local gradients of the material flow
(Fig.~\ref{fig:Fig2}; Supplementary Section IV).

This model makes the immediately testable prediction that the separation speed $v_\text{sep} = |\dot{\bX}(\uvc{s})-\dot{\bX}(-\uvc{s})|$ between oppositely directed and aligned segments of sperm is independent of the material's local polarity $\bP(\bx,t)$ (Supplementary Section IV). In contrast to $\dot{\bx}$, which includes a contribution from the background material flow $\bU$, $v_\text{sep}$ depends only on the sperm's propulsion speed $v_s$. Using data from photobleaching experiments (Fig.~\ref{fig:Fig3}), we measured the speed difference between oppositely propagating bleached sperm segments as a function of the splitting distance $\Lambda$, a proxy for the local material polarity. We found that $v_\text{sep}$ is largely uncorrelated with the local polarity of the material, in agreement with DER simulations of various arrangements of co- and counter-propagating active filaments in planar assemblies (Fig.~\ref{fig:Fig4}\textbf{d}, Extended Data Fig.~\ref{fig:EDF6}; Supplementary Section IV).

\vspace*{0.5em}
\noindent \textbf{A micro-macro theory for material flows}
\vspace*{0.2em}

\noindent Building on these findings, we now investigate the emergence of the material flow $\bU$ through a coarse-grained momentum balance equation:
\begin{equation}
    -\eta \rho \bU + \nabla \cdot \bSigma = \mathbf{0}.
\end{equation}
Here, $\eta$ is a friction coefficient and $\bSigma$ is the coarse-grained stress tensor in the flagellar material that has four contributions $\bSigma = \bSigma^\text{a} + \bSigma^\text{el} + \bSigma^\text{st} - \Pi \bI$ (Extended Data Fig.~\ref{fig:EDF5}). Notably, we find that the active stress $\bSigma^\text{a}$, which emerges naturally from equal and opposite frictional and contact propulsion forces, is extensile and given by $\bSigma^\text{a} = \alpha \rho^2 (\bP \bP - \bQ)$, where $\alpha > 0$ is a measure of activity and $\mathbf{Q}(\bx) = \rho^{-1} \int_{\mathbb{S}^2} \bq \bq \psi(\bx)  \md \bq$ is the nematic tensor of the material. The $\rho^2$ dependence reflects the active stress's origins in steric interactions between counter-propagating waves. Our micro-macro framework directly captures the polarity dependence of this stress, highlighting that when the material is polar ($|\bP| \sim 1$) and sperm segments are locally immotile, then $\bSigma^\text{a} \approx \mathbf{0}$. That $\alpha>0$ marks the active stress as extensile, as was revealed by the direction of motion of $+1/2$-order defects (Fig.~\ref{fig:Fig2}\textbf{g}; Supplementary Video~2) \cite{copenhagen2021topological}. 
The remaining stresses account for appropriate long-wavelength dynamics of elastic filaments ($\bSigma^{\text{el}}$) \cite{de1993physics}, aligning steric interactions among the flagellar backbones ($\bSigma^{\text{st}}$), and incompressibility of the material flows via a Lagrange multiplier ($\Pi$) (Supplementary Section IV). 

A linear stability analysis around an apolar aligned state of the material predicts an instability reminiscent of Euler buckling of elastic filaments \cite{audoly2000elasticity} that differs from the generic bend-instability of active suspensions in a Stokesian fluid \cite{aditi2002hydrodynamic,saintillan2008instabilities1} (Supplementary Section IV).

To probe the nonlinear dynamics of this system, we directly evolve the polarity and nematic fields, $\bP(\bx,t)$ and $\bQ(\bx,t)$, respectively, using a well-tested closure approximation under which the distribution function can be reconstructed \cite{weady2022thermodynamically}(Supplementary Section IV). Although the model is inherently 3D, simulations were performed in a disk, as the main features of the system's dynamics and organization are discernible in 2D and unaltered when 3D dynamics are experimentally suppressed (Fig.~\ref{fig:Fig5}\textbf{a},\textbf{b}; Extended Data Fig.~\ref{fig:EDF4}; Supplementary Video~5). 

Time evolution of the director field reveals a simulated material in a highly aligned state whose material velocity $\bU$ is characterized by the propulsion of $+1/2$-order defects. We also observe the spontaneous emergence and annihilation of $+1/2$- and $-1/2$-order defect pairs in the bulk which, in a disk, preserves a topological charge of $+1$ (Fig.~\ref{fig:Fig5}\textbf{c}; Supplementary Video~9). Simulations also provide a direct window into the evolution of the polarity field: We find that while the material is largely apolar, localized polar regions appear and disappear over time (Fig.~\ref{fig:Fig5}\textbf{e}), and their appearance correlates with the emergence of $+1/2$-order defects in the director field. For the chosen parameters (Supplementary Section IV), the model also reproduces experimentally observed correlations and statistics such as the separation of scales between defect and swimming speed, statistics of polar regions, and the correlations between material vorticity and defect speeds  (Extended Data Fig.~\ref{fig:EDF7}).

The present micro-macro framework of the coarse-grained model builds on the sperm's translational flux. Unique to this approach is now our ability to probe the microscopic organization of the flagellar material and explore the model's prediction for flagellar dynamics. To this end, we reconstruct $\psi(\bx,\bq,t)$ (with $\bq = (\cos \theta, \sin \theta)$ in 2D) from $\bP(\bx,t)$ and $\bQ(\bx,t)$ in three distinct regions of the material (Fig.~\ref{fig:Fig5}\textbf{f}; Supplementary Video~9, Supplementary Section IV) \cite{weady2022thermodynamically}. We find that in the vicinity of a $-1/2$-order defect, the orientational distribution of the microstructure is roughly isotropic, with $S\approx 0$, compared to the aligned regions where $S \approx 1$. In a nematic polar region, the distribution peaks predominantly around one angle, while in a nematic apolar region, it peaks around two angles separated by $\pi$, corresponding to opposite orientations of the local director field. We demonstrate how this underlying microstructure influences the motion of individual sperm segments through numerical bleaching experiments (Fig.~\ref{fig:Fig5}\textbf{g}). In nematic regions, sperm segments split into two counter-propagating bands along the local director, with their relative density determined by the local polarity. Thus, in apolar regions, counter-propagating bands have equal densities, capturing the behavior seen experimentally in Fig.~\ref{fig:Fig3}\textbf{c}, while in the polar areas, segments move unidirectionally, resembling the scenario observed in Fig.~\ref{fig:Fig3}\textbf{e}. In contrast with bleached segments in nematic regions, bleached segments near a $-1/2$-order defect split isotropically, capturing the behavior observed experimentally when photobleaching sperm in a tri-junctional arrangement (Fig.~\ref{fig:Fig3}\textbf{g}). 
Using the translational flux from our theory, the reconstructed distribution function at the scale of the entire material further allows us to infer the relative motion of marked sperm segments in the mass (Fig.~\ref{fig:Fig5}\textbf{d}).

\vspace*{2mm}
\vspace*{0.5em}
\noindent \textbf{Discussion}
\vspace*{0.2em}

Sperm, nearly universal for sexual reproduction in the animal and plant kingdoms, are among the most morphologically diverse cell types \cite{Fitzpatrick2022}. Still, the nature and consequences of sperm activity have been intensely studied for only a canonical few, such as mammalian and sea urchin sperm. \textit{Drosophila}'s ultralong sperm, which arise from intense post-copulatory sexual selection \cite{Miller2002}, is a striking beyond-this-canon example. Our work 
establishes such giant sperm {\it in vivo} as a novel and developmentally relevant active matter system, whose extensile stresses arise from sperm's rapid and collective counter-propagation through densely-packed storage organs. We propose that the sperm's aligned, unentangled state, critical to reproductive success, is maintained by these self-same stresses. Future studies of beyond-the-canon sperm will uncover an expanding repertoire of surprising behaviors, and inspire fundamental questions at the intersection of biological active matter and evolution.

Actively reptating sperm assemblies are not only found in the \textit{D. melanogaster} male: Females can also maintain sperm for prolonged periods ($\sim$2 weeks) in dedicated storage organs \cite{BlochQazi2003,Gao2003,Neubaum1999}\textemdash as documented in numerous other species (Fig.~1 in \cite{Holt2016SpermSI}). Approximately $80\%$ of stored sperm in the female are found in the seminal vesicle (SR), a narrow ($d \sim 10-30 \ \mu$m) and long ($L \sim 2$-mm) blind-ended tube \cite{Kaufman1942,Lefevre1962,BlochQazi2003,Gao2003,Neubaum1999, Manier2010}. We examined the storage organs from females mated with males that produce fluorescently-labeled sperm and found that, consistent with previous reports (Supplementary Video S1 in \cite{Manier2010}), sperm in the SR are persistently motile (Extended Data Fig.~\ref{fig:EDF8}; Supplementary Video~10). Our live imaging of their flagella further demonstrated that, as in the SV, sperm in the female SRs are densely packed and locally aligned, 
and again exhibit continuously evolving spontaneous material flows with slow-moving defects and fast laning individuals (Extended Data Fig.~\ref{fig:EDF8}; Supplementary Video~10). Notably, $\sim$$50\%$ of stored sperm fertilize mature oocytes \cite{Kaufman1942, Lefevre1962}, raising questions about how sperm are extracted, almost singly, from this state.


Active matter concepts are emerging as a powerful lens through which to study biological phenomena at the microscale \cite{needleman2017active}. Non-equilibrium dynamics driven by active extensile stresses drive collective motion in bacterial suspensions \cite{saintillan2008instabilities1}, help shape the mitotic spindle \cite{brugues2014physical}, may fluidize developing epithelial tissues \cite{aditi2002hydrodynamic}, and are conjectured to drive large-scale nuclear chromatin fluctuations and the unfolding of long crumpled active polymers and their organization into nematic alignment (Fig. 3 and Fig. S1 in \cite{saintillan2018extensile}) \cite{mahajan2022euchromatin}. This work proposes their functional relevance to reproduction in a model organism through a novel mechanism. Whereas jostled passive elastic filaments inevitably entangle under confinement \cite{Belmonte2001,Raymer2007}, ultralong motile sperm in the animal's storage organs remain unentangled despite their dense packing and long-term storage, possibly through self-generated stresses stemming from their persistent active reptation. The observed active fluidization inside the SV is reminiscent of the dynamics observed in glassy active systems \cite{bi2016motility,mandal2020extreme} and in epithelial tissues \cite{rozman2024cell}
, suggesting a generalizable role for extensile activity in maintaining dense living matter unjammed and functional. Given the inherent polarity of individual sperm, the emergent apolarity of the material is also striking -- presumably arising from dynamic reversals in the direction of the bending waves of individual sperm \cite{Yang2011, Mojica1996}, and local topological rearrangements driven by the interlacing of the filaments' free ends with neighboring sperm.

\vspace*{0.5em}
\noindent \textbf{Acknowledgments}
\vspace*{0.2em}

\noindent The authors thank the NYULH DART Microscopy Lab—Dr. Alice Liang and Joseph Sall—for assistance with electron microscopy, supported by Cancer Center Grant P30CA016087; the Gemini300SEM with 3View was purchased with NIH S10 OD019974. B.C. acknowledges support from the Department of Atomic Energy, Government of India (project no. RTI4001). The authors are grateful to Kurt Fan and Sophia Zhang for help with sperm tracking from SBF-SEM data; Jonathan Jackson (MPI-CG) and Nicolas Romeo (U. Chicago) for image analysis assistance; and Scott Weady, Reza Farhadifar, Lisa Brown (all Flatiron Institute), and Frank Jülicher (MPI-CS) for valuable discussions. Computations were carried out at the Flatiron Institute’s Scientific Computing Core. Confocal microscopy data were acquired at the CCBScope Observatory.


\vspace*{0.5em}
\noindent \textbf{Author contributions}
\vspace*{0.2em}

\noindent JIA conceived the project. JIA designed and conducted experiments with input from BC and MJS. BC and MJS developed the theoretical model and implemented the continuum simulations. BP performed the DER simulations and developed the data visualization framework for Fig.~\ref{fig:Fig1}. JIA and BC performed image and data analysis. JIA and BC wrote the original manuscript with the input of all authors. All authors reviewed the manuscript.

\vspace*{0.5em}
\noindent \textbf{Data availability}
\vspace*{0.2em}

\noindent All the data that support the plots within this paper and other findings of this study are available from the corresponding author upon reasonable request.

\vspace*{0.5em}
\noindent \textbf{Code availability}
\vspace*{0.2em}

\noindent The computational methods that support the plots within this paper are
described in the Supplementary Information and the underlying code
is available from the corresponding author upon reasonable request.

\vspace*{0.5em}
\noindent \textbf{Ethics declarations}
\vspace*{0.2em}

\noindent The authors declare no competing interests.

\let\oldaddcontentsline\addcontentsline
\renewcommand{\addcontentsline}[3]{}
\bibliography{bibliography}
\let\addcontentsline\oldaddcontentsline

\newpage

\onecolumngrid
\vspace{1em}
\begin{center}
\section*{Extended Data Figures}
\end{center}
\vspace{1em}

\begin{extendedfigure}[b]
  \centering
  \includegraphics[width=0.8\textwidth]{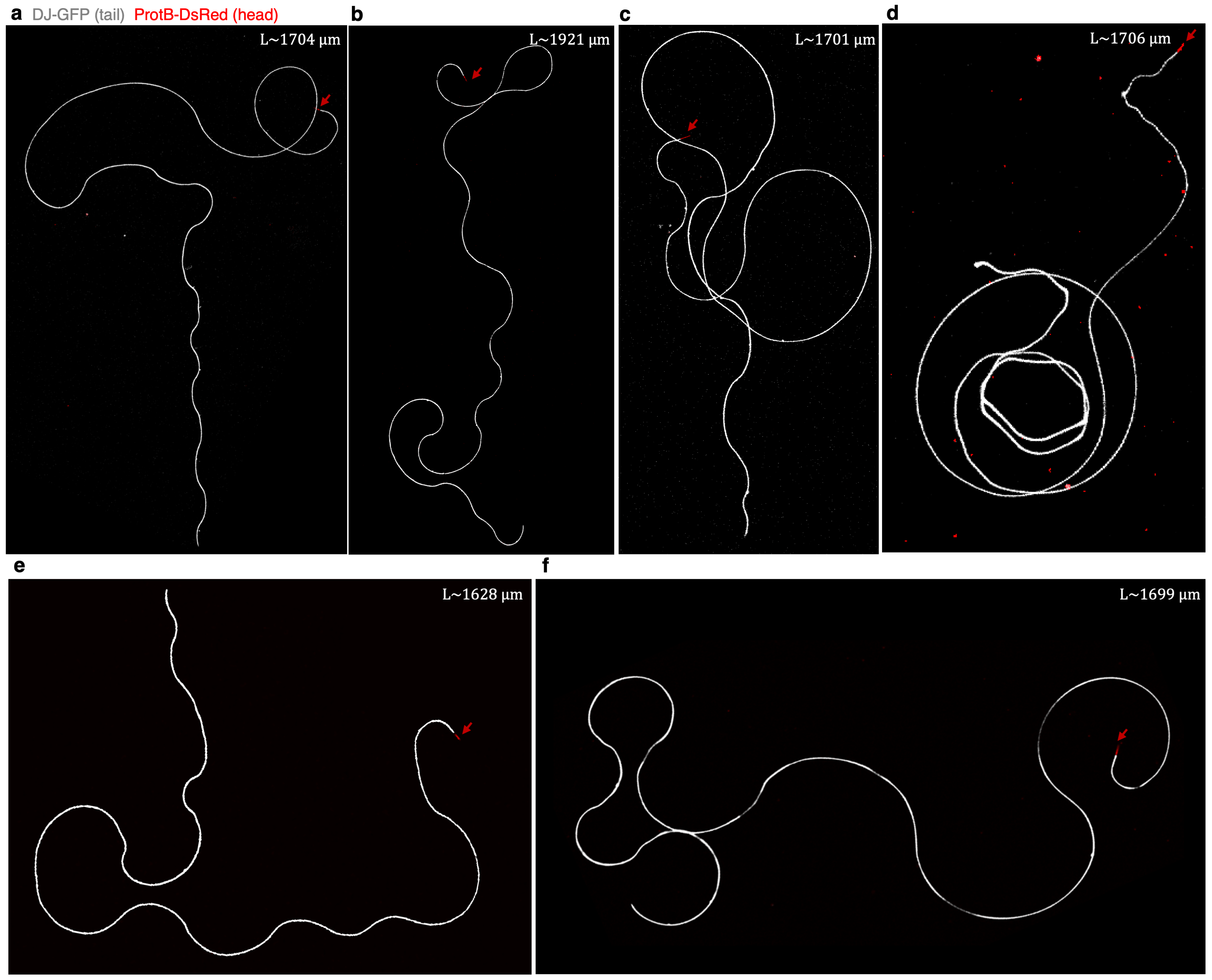}
  \caption{\textbf{Confocal microscopy images of individual \textit{Drosophila melanogaster} sperm.} \textbf{a-f}: Examples of individual sperm with fluorescently labeled flagella (or tails) (DJ-GFP) and heads (ProtB-DsRed), extracted from the male's storage organ, the seminal vesicle (SV). Sperm lengths, in $\mu$m, are indicated; red arrow points to each of their heads. These results are consistent with those reported in \cite{Joly1994}, where sperm lengths were estimated by measuring the lengths of individual sperm and of sperm within mature cysts in the testes. The sperm shown in \textbf{f} was used to generate the outline of the sperm shown in Fig.~1\textbf{a} of the main text.}  
  \label{fig:EDF1}
\end{extendedfigure}

\twocolumngrid

\begin{extendedfigure*}
\centering
\includegraphics[width=0.8\linewidth]{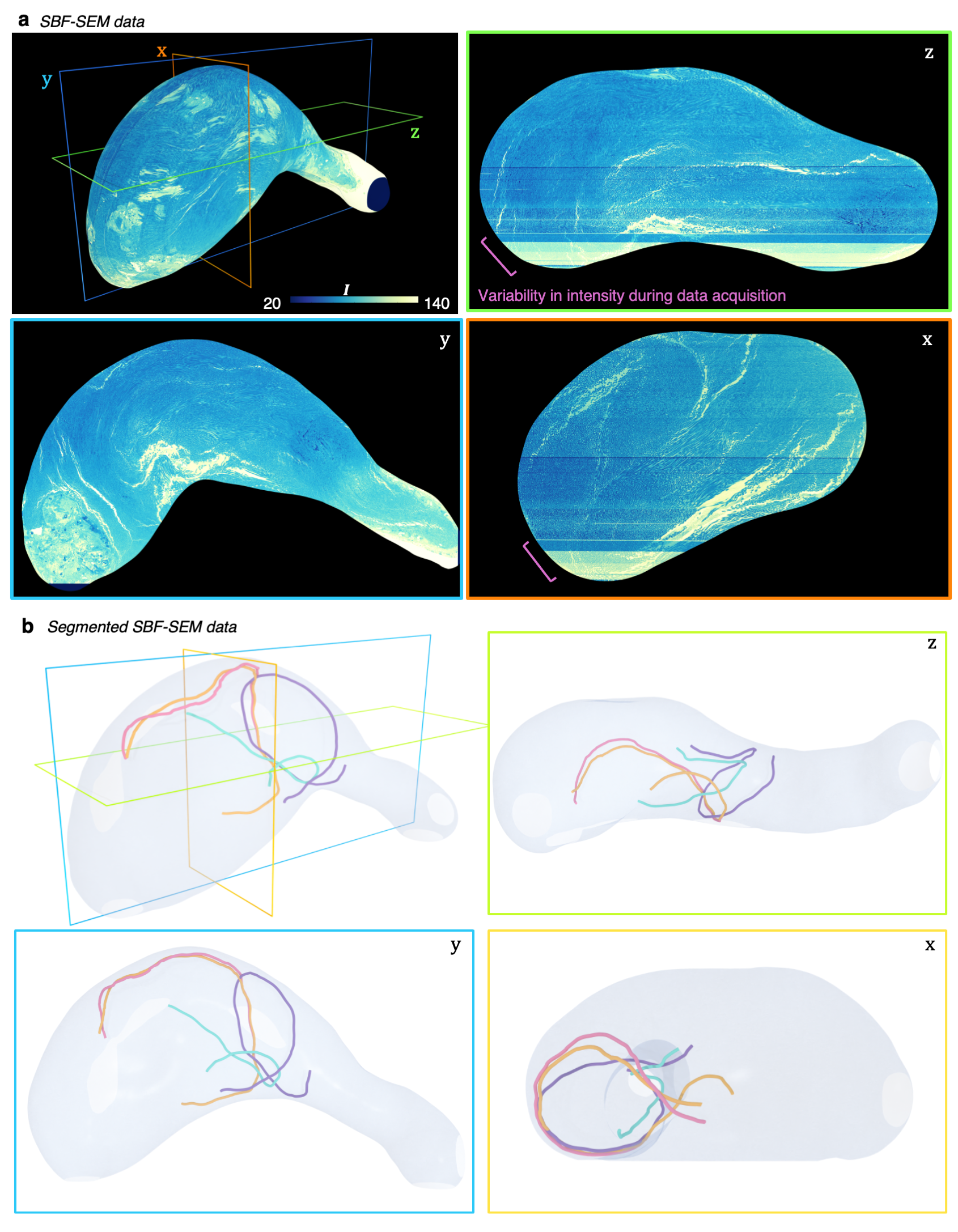}
\caption{\textbf{Reconstructions of sperm from serial block-face scanning electron microscopy (SBF-SEM) data.} \textbf{a,} A 3D rendered image of sperm in the SV, colored by intensity $I$, acquired through SBF-SEM (Supplementary Video~2). Different cross-sectional views are shown, demonstrating the dense and aligned packing of sperm throughout the SV. The observed variation in the intensity of the data arises from variability in the microscope and camera settings during data acquisition. \textbf{b,} 3D rendered image of the reconstructed SV showing four partially segmented sperm based on SBF-SEM data (Fig.~S3). Sperm were segmented manually by tracking and highlighting the sperm’s cross sectional area in successive slices using Dragonfly 4.1. Figs.~1\textbf{d,e} in the main text show these segmented sperm overlaid with the SBF-SEM data of the entire SV. \textbf{c,} Different views of the SV and the four sperm (color-coded as in \textbf{b}). The visualizations were generated in ParaView.}
\label{fig:EDF2}
\end{extendedfigure*}

\begin{extendedfigure*}
\centering
\includegraphics[width=0.9\linewidth]{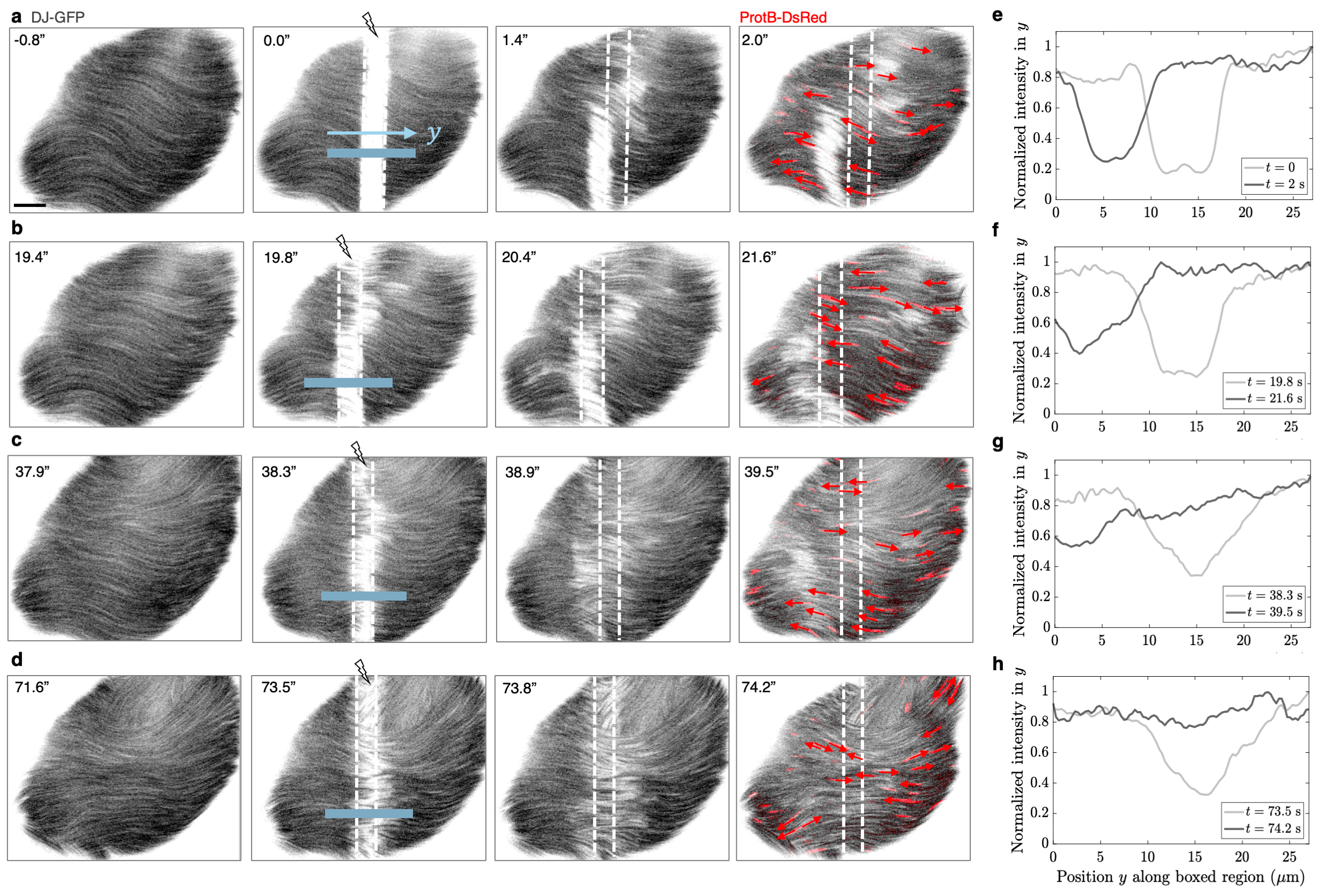}

\caption{\textbf{Photobleaching experiments reveal the evolution of the material's polarity field and organization.} \textbf{a-d} Snapshots from a time lapse of fluorescently labeled sperm (DJ-GFP, labels flagella; ProtB-DsRed, labels heads) in the SV, showing recovery of fluorescence and evolution of the material’s polarity after four successive photobleaching events ($t = 0$ s, $t = 19.8$ s, $t = 38.3$ s, $t = 73.5$ s). A bleach line, transverse to the sperm’s nematic director, is demarcated by dashed white lines. Sperm heads and their direction of propagation (arrows) are shown in the final time point of each bleach event. \textbf{e-h}, Normalized intensity as a function of distance along the highlighted band (blue) at bleaching and at a subsequent time point. As seen from the timelapses and the adjoining quantification, the material’s polarity evolves. Scale bar = $10$ $\mu$m.}
\label{fig:EDF3}
\end{extendedfigure*}

\begin{extendedfigure*}
\centering
\includegraphics[width=0.8\linewidth]{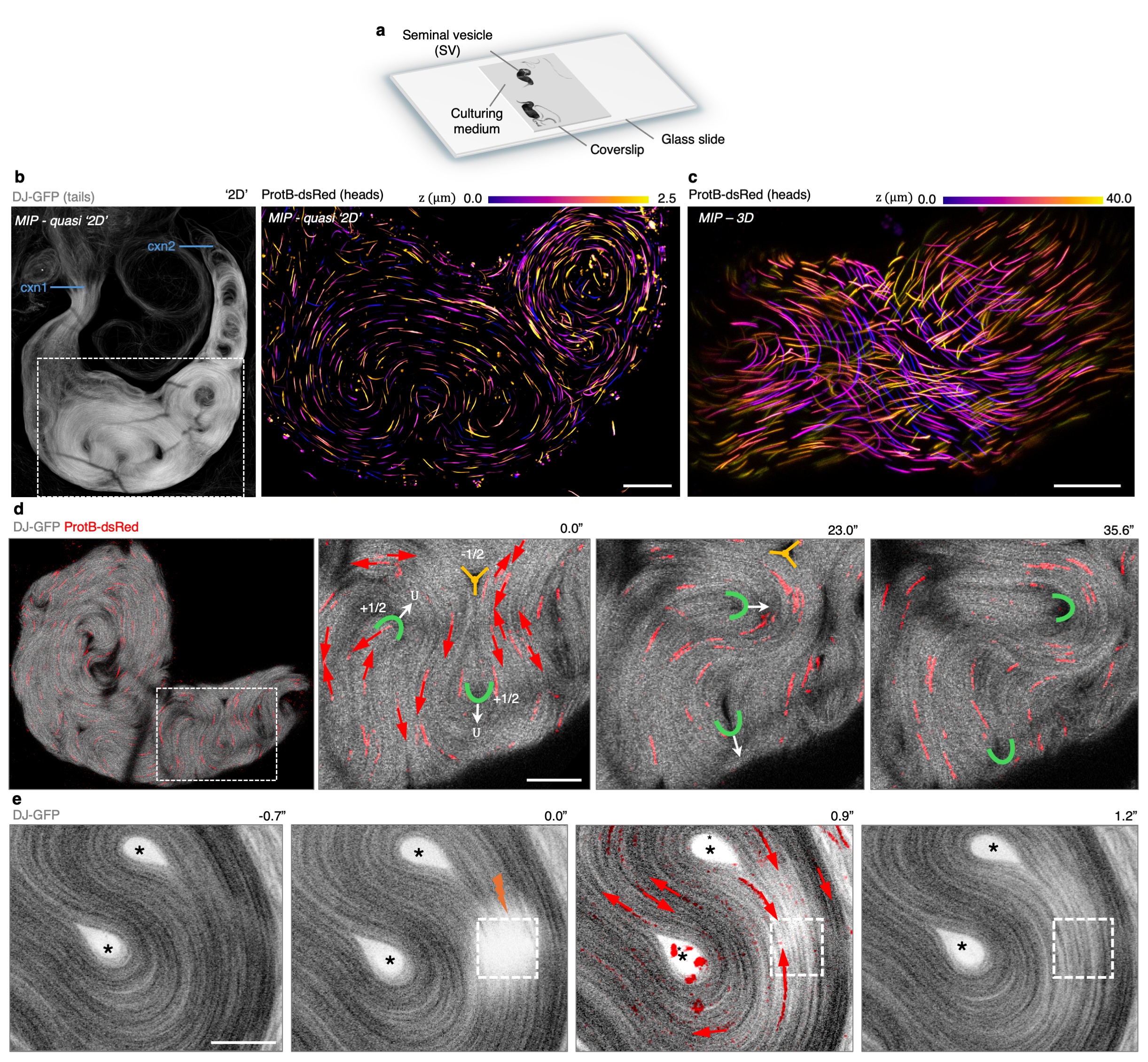}
\caption{\textbf{Main features of the 3D spatiotemporal dynamics are largely preserved in compressed quasi-2D samples.} \textbf{a,} Schematic of the experimental setup. Dissected SVs in Schneider’s medium were aspirated using a pipette with a cut tip, placed on a glass slide, and compressed with a coverslip that was partially sealed (two sides) with clear nail polish for live imaging. \textbf{b,} \textit{(Left)} Confocal image of fluorescently labeled sperm (DJ-GFP) in a SV disconnected from the testis (cxn1) and the ejaculatory duct (cxn2) under a coverslip. In this experimental setup, the SV is compressed, causing rapid exit of sperm from the severed ends; the remaining sperm are confined to a $\sim$$3-5$ $\mu$m depth, which is why this configuration is referred to as quasi-2D. \textit{(Right)} Maximum intensity projection (MIP) of sperm heads in the boxed region in \textbf{a}, demonstrating the absence of variation in the material’s director field in $z$ – as opposed to the 3D case of an uncompressed SV shown in \textbf{c}, and in Figs.~1\textbf{d,f} in the main text. \textbf{c,} 3D-rendered image of a $z$-stack of an uncompressed SV in a glass-bottom imaging dish, highlighting the varying orientations of sperm heads, a proxy for material organization, as a function of depth in the SV (color encodes position in $z$ in $\mu$m). \textbf{d,} Snapshots from a time series of the boxed region of a compressed SV, highlighting two $+1/2$- and $-1/2$-order defects. Arrows overlaid on the sperm heads (ProtB-DsRed) indicate the direction of propagation of individual sperm. Scale bars in \textbf{b}-\textbf{c} = 20 $\mu$m. \textbf{e,} Snapshots from a time lapse in compressed SV, in which a rectangular region (10 $\mu$m x 10 $\mu$m) in the SV was photobleached at $t=0$ s. Subsequent time frames show the counter-propagation of sperm and eventual recovery of fluorescence through the influx of counter-propagating sperm. Stars indicate the positions of +1/2-order defects that are devoid of sperm. Scale bar = $10$ $\mu$m.}
\label{fig:EDF4}
\end{extendedfigure*}

\begin{extendedfigure*}
\centering
\includegraphics[width=0.7\linewidth]{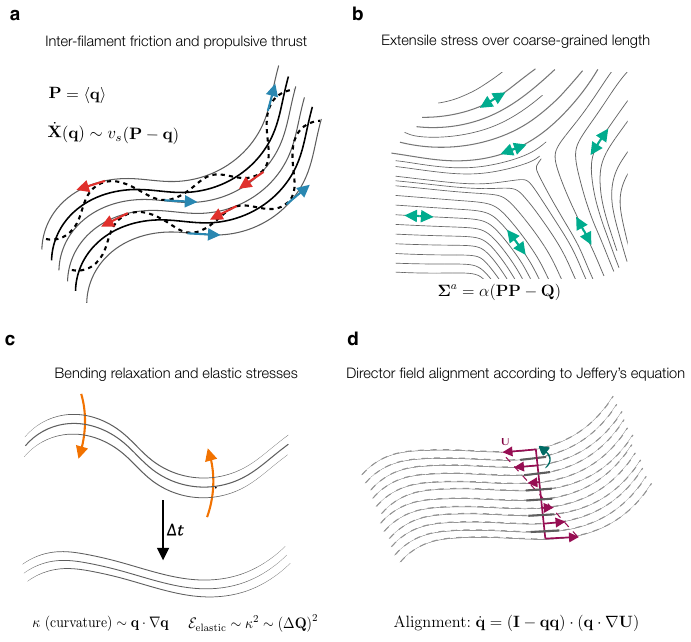}
\caption{\textbf{Schematics of the mechanical interactions that form the basis of the micro-macro theory.} \textbf{a,} Inter-filament friction (red arrows) and thrust (blue arrows) forces resulting from mechanical interaction between bending waves (dashed lines) in the sperm assembly. The net propulsive thrust and, in turn, $\dot{\bX}(\bq)$, depend on the local polarity $\bP(\bx)$ of the material. The model neglects the role of possible phase-synchronization in neighboring waves. \textbf{b,} The contact interactions that give rise to directed motility also lead to extensile stresses over length scales much larger than those over which contact interactions occur. These dipolar stresses (double-headed arrows) depend on the nematic tensor $\bQ$ and the polarity field $\bP$. \textbf{c,} Field theoretical description of elasticity in aligned fibers that mimic the passive bending relaxation of sperm segments. \textbf{d,} Alignment and rotation of sperm segments with the mean-field material flow $\bU$ (portrayed as a shear flow using purple arrows). The green arrow indicates the rotation of sperm segments.}
\label{fig:EDF5}
\end{extendedfigure*}

\begin{extendedfigure*}
\centering
\includegraphics[width=0.6\linewidth]{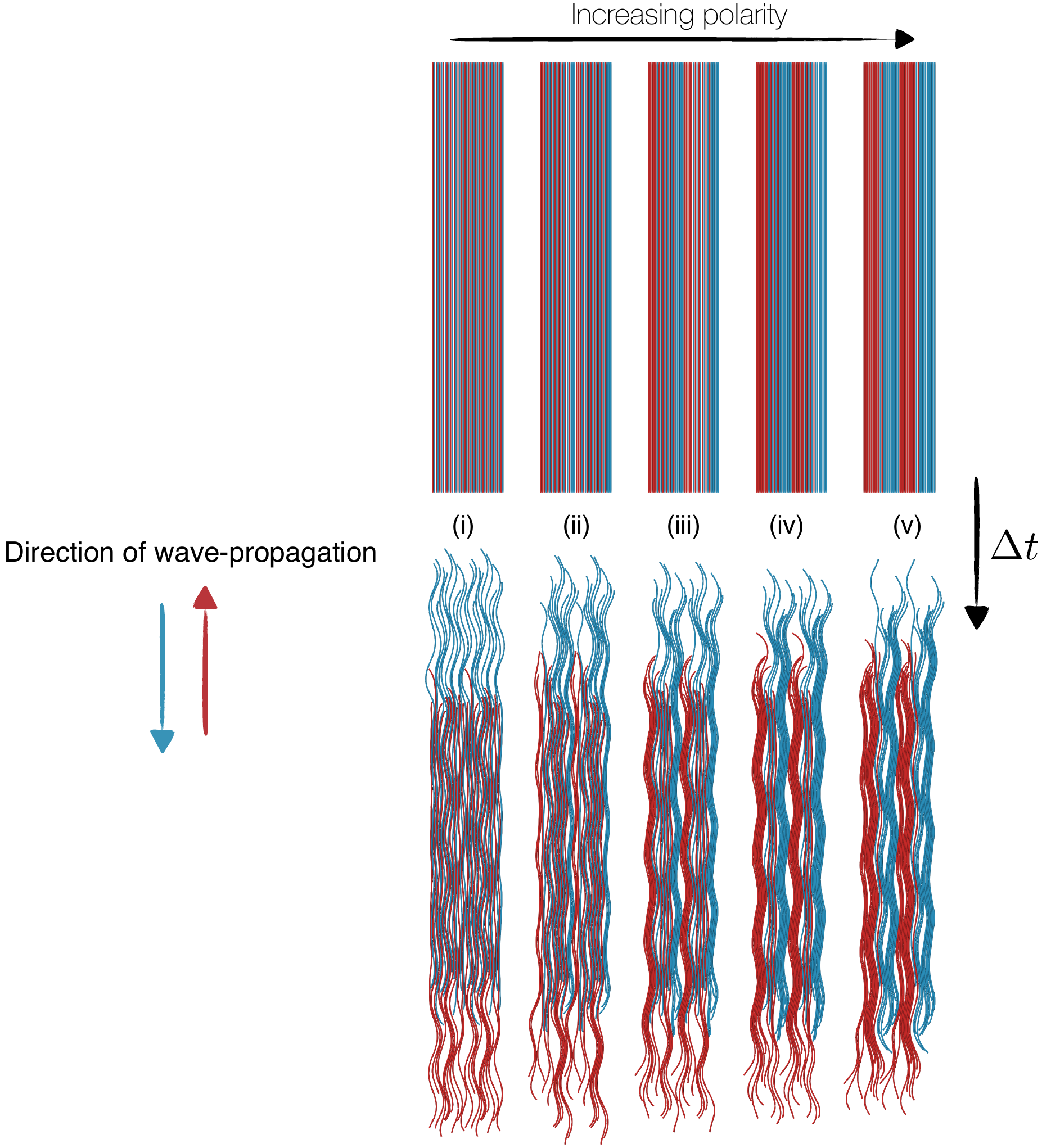}
\caption{\textbf{Discrete Elastic Rod (DER) simulations of active filaments in various interleaved configurations.} Active filaments are interleaved to various extents. The polarity of these configurations increases from left to right (top row arrow). The polar configurations can be associated with coherently moving groups of sperm with high $\Lambda$ in the experiments. The bottom row highlights the splitting dynamics of the configurations at a later time point. 
The aspect ratio $300$ filaments are modeled using $N=300$ weakly extensible segments with linear elastic stretching (along each segment) and bending rigidity (between adjacent segments) comparable to that of a bull sperm flagella \cite{rikmenspoel_bull_sperm_1984}. Undulatory motion is driven by a time-varying sinusoidal rest curvature with five periods of osculation long each filament and a random phase shift per filament to desynchronize the initial state. The evolution of the filaments is governed by isotropic passive drag from the suspending fluid, chosen such that isolated filaments do not self-propel. Each of these five simulations contains 20 filaments\textemdash 10 propagating waves downward and 10 propagating waves upward\textemdash at an area fraction of 80\% with periodic boundary conditions transverse to the long axis of the filaments.}
\label{fig:EDF6}
\end{extendedfigure*}

\begin{extendedfigure*}
\centering
\includegraphics[width=0.7\linewidth]{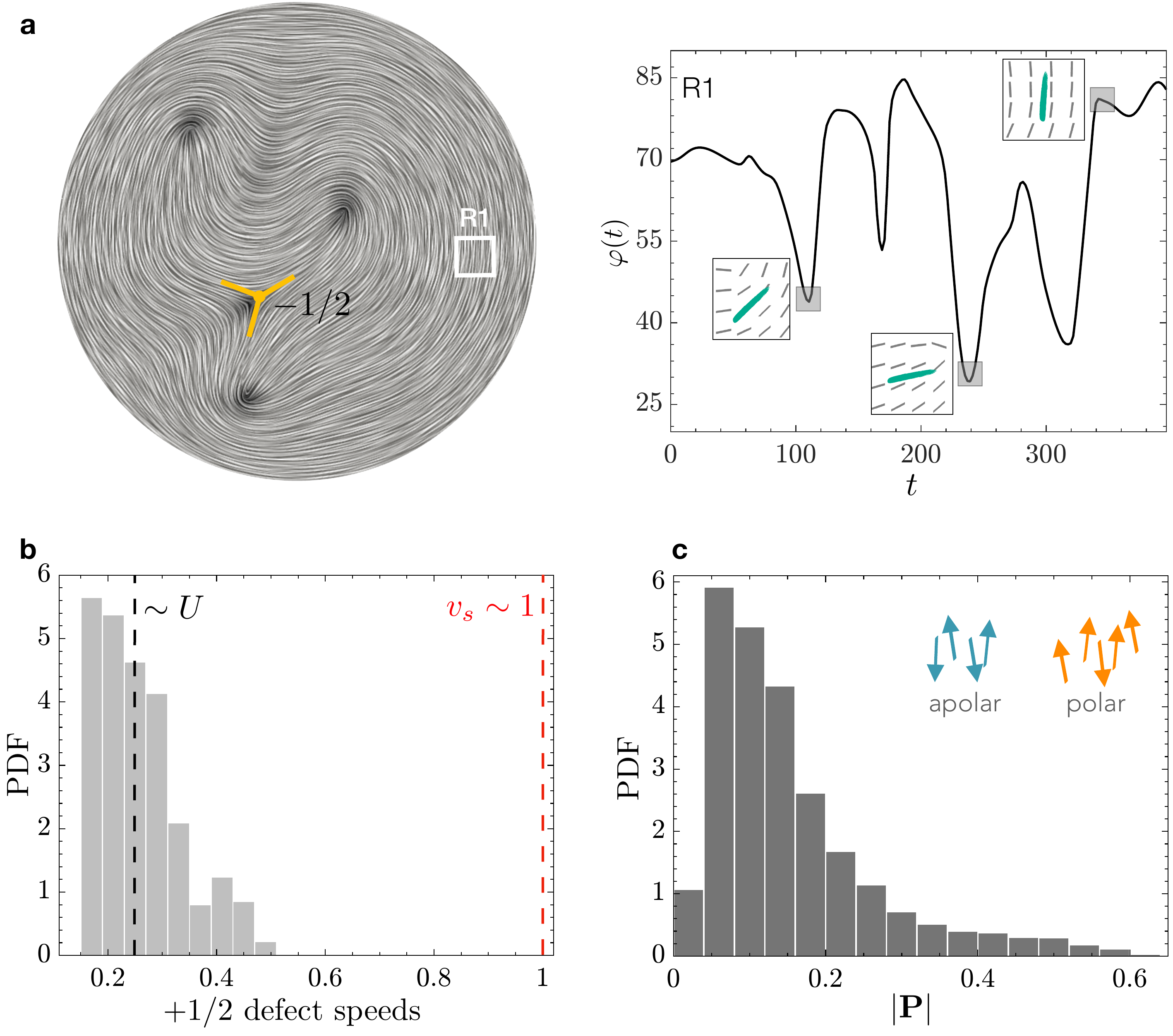}
\caption{\textbf{Director orientation dynamics and statistics of the material velocity $U$ and polarity $P$ from continuum simulations.} \textbf{a,} \textit{Left}: Snapshot of the director field of the flagellar material from simulations of the hydrodynamic model in a disk, displayed using Line Integral Convolution (LIC), where the color encodes the strength of the nematic order parameter $S$. A -1/2-order defect is highlighted in yellow. \textit{Right:} Plot of the evolution of the local director orientation in R1 (white box); insets show the spatial director organization. We estimate a dimensionless time scale $\tau \sim 30$ for the organization of the material and an associated velocity scale $U \sim R/\tau \sim U_\text{defects} \sim 0.3$, where $R$ is the system size. \textbf{b,} Histogram of the distribution of the defect speeds $U_\text{defects}$, estimated by tracking $+1/2$-order defects. For the chosen parameters (Supplementary Section IV), the swimming speed of flagellar segments $v_s \sim 1$ is separated in scale from the average material speed, as observed in experiments (Fig.~2\textbf{g} in the main text, Supplementary Video 3). This direct measurement of defect speed is in agreement with the scaling estimate obtained using the material vorticity. \textbf{c,} Histogram of $|\bP|$ sampled across the simulation domain, demonstrating the largely apolar nature of the material. Schematized arrows reflect the microstructural organization of sperm segments in qualitatively different regions of $|\bP|$.}
\label{fig:EDF7}
\end{extendedfigure*}

\begin{extendedfigure*}
\centering
\includegraphics[width=0.7\linewidth]{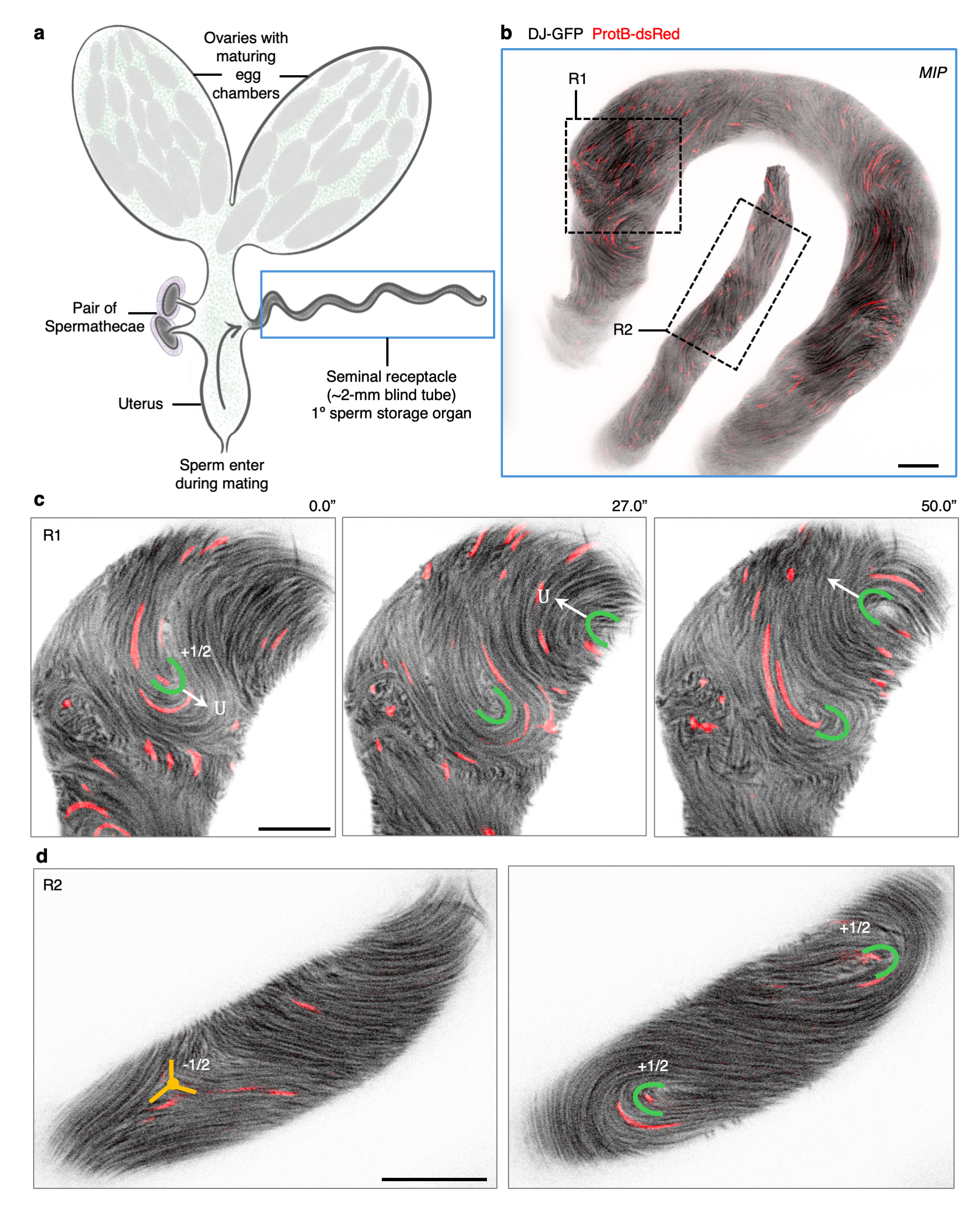}
\caption{\textbf{Live imaging experiments reveal similar organization and dynamics of sperm in the female's primary sperm storage organ as in the male.} \textbf{a,} Simplified schematic of the \textit{Drosophila melanogaster} female reproductive system. A pair of ovaries houses developing egg chambers, which ultimately give rise to the oocyte (egg cell). During mating, sperm deposited by the male find their way to the sperm storage organs (SSO), where they remain until fertilization for up to two weeks. The primary SSO is a blind $\sim$2-mm tube called the seminal receptacle (boxed), which houses most ($\sim$$80 \%$) of the transferred sperm \cite{Lefevre1962}; the remainder are stored in a pair of mushroom-shaped organs called the spermatheca. \textbf{b,} MIP ($\sim 40 \ \mu$m deep) of stored sperm with fluorescently labeled heads (ProtB-DsRed) and tails (DJ-GFP) in the seminal receptacle. Snapshots of optical sections from time-lapses of the boxed regions are shown in \textbf{c} (R1) and \textbf{d} (R2). \textbf{d,} Snapshots from a time-lapse highlighting a pair of $+1/2$-order defects moving in opposing directions with speed $U$ in the wider (posterior) region of the seminal receptacle (diameter $\sim 30 \ \mu$m). \textbf{e,} Snapshots from a time-lapse showing the emergence of $+1/2$- and $-1/2$-order defects in the narrower (anterior) region (diameter $\sim 10 \ \mu$m) of the seminal receptacle. As in the male’s SV, sperm’s flagella in the female’s seminal receptacle are highly aligned, resembling a continuous nematic material with spontaneously emerging and slowly moving defects. Individual sperm, in contrast, counter-propagate within lanes formed by the flagellar backbone. Scale bars in \textbf{b}-\textbf{d} = 10 $\mu$m.}
\label{fig:EDF8}
\end{extendedfigure*}

\end{document}